\documentclass[%
 aip,
 amsmath,amssymb,
 reprint,%
]{revtex4-2}

\usepackage{graphicx}% Include figure files
\usepackage{dcolumn}% Align table columns on decimal point
\usepackage{bm}% bold math
\usepackage[utf8]{inputenc}
\usepackage[T1]{fontenc}
\usepackage{mathptmx}
\usepackage{etoolbox}
\usepackage{multirow}
\usepackage{booktabs}

\makeatletter
\def\@email#1#2{%
 \endgroup
 \patchcmd{\titleblock@produce}
  {\frontmatter@RRAPformat}
  {\frontmatter@RRAPformat{\produce@RRAP{*#1\href{mailto:#2}{#2}}}\frontmatter@RRAPformat}
  {}{}
}%
\makeatother
\DeclareMathOperator*{\argmin}{arg\,min}
\begin{document}

\preprint{AIP/123-QED}

\title{Analytic nuclear gradients for electronic resonances with the equation-of-motion coupled-cluster method and complex basis functions}

\author{Robin E. Moorby}
\email{robin.moorby@kuleuven.be}
\affiliation{Department of Chemistry, KU Leuven, Celestijnenlaan 200F, 3001 Leuven, Belgium}

\author{Thomas-C. Jagau}
\email{thomas.jagau@kuleuven.be}
\affiliation{Department of Chemistry, KU Leuven, Celestijnenlaan 200F, 3001 Leuven, Belgium}

\date{\today}

\begin{abstract}
We present the general theory for and an efficient implementation of the analytic nuclear gradient for states which are metastable with respect to electronic decay within the equation-of-motion coupled-cluster singles-and-doubles family of methods using complex-scaled basis functions. We highlight the differences of our approach with the complex-absorbing potential method and outline a computational protocol for geometry optimisations and we use the method to compute equilibrium structures and harmonic vibrational frequencies of a number of temporary anion shape resonances and core-ionised Feshbach resonances.
\end{abstract}

\maketitle

Electronic resonances are states embedded in the continuum that are unstable with respect to electronic decay; examples include temporary anions,\cite{Simons:Anions:2011,Herbert:Anions:2015,Alizadeh:Anions:2015} core-ionised and core-excited states,\cite{Agarwal:Xrays:2013,Norman:Xrays:2018,Zimmermann:Xrays:2020} autoionising Rydberg states,\cite{Platzman:Rydberg:1962, Klinker:Rydberg:2018, Plunkett:Rydberg:2019, Rabadan:Rydberg:2021} Stark resonances,\cite{Scrinzi:Stark:2006,Gallmann:Stark:2012} and intermolecular Coulombic decay.\cite{Cederbaum:ICD:1997,Ouchi:ICD:2011,Jahnke:ICD:2020} These states are involved in various chemical reactions including dissociative electron detachment,\cite{Fabrikant:DEA:2017,Ingolfsson:DEA:2019} dissociative recombination\cite{Bardsley:DissociativeRecombination:1968,Brian:DissociativeRecombination:1990} and Coulomb explosion\cite{Ouchi:ICD:2011,Jahnke:ICD:2020} and are relevant to numerous processes such as plasmonic catalysis\cite{Aslam:Plasmon:2018,Martirez:Plasmon:2021}, radiation damage\cite{Boudaiffa:Damage:2000,Alizadeh:Anions:2015} and astrochemistry.\cite{Millar:Astro:2017}

Resonance wavefunctions are not square-integrable meaning that they are not amenable to bound-state quantum chemical methods. Besides approaches based on scattering theory,\cite{Gorfinkiel:Review:2020} an elegant solution to this problem lies in the family of complex-variable methods\cite{Moiseyev:Book:11,Jagau:Review:2017,Jagau:Review:2022} which includes full and exterior complex scaling,\cite{Aguilar:CS:1971,Balslev:CS:1971,Simon:CS:1972,Moiseyev:CSH2:1979,Moiseyev:CSReview:1998,Bravaya:ComplexScaling:2013} complex absorbing potentials (CAPs),\cite{Austin:CAP:1985,Meyer:CAP:1993,Zuev:CAP:14,Jagau:CAP:13,Jagau:PES:14} and complex basis functions (CBFs).\cite{McCurdy:BO:1978,White:CBF:2015,White:CBFSCF:2016,White:CBFCCSD:2017} An advantage of the CAP approach, where an imaginary potential is added to the Hamiltonian, is that if a software can evaluate the one-electron CAP integrals and handle complex-valued density matrices, standard bound state routines can be reused without major changes. Also, in the projected CAP approach, the CAP equations are solved \emph{a posteriori} to a normal electronic structure calculation in a basis of pseudocontinuum states, which enables substantial computational savings.\cite{Sommerfeld:CAPCI:2001,Santra:CAP:2001,Gayvert:CAP:2022}

CBFs on the other hand require reimplementation of all Hamiltonian and property matrix elements over non-standard basis functions, and thus implementations are generally more challenging. However, the method holds several advantages over the CAP method, both formal and practical. %Perhaps most importantly, CBFs can be applied to all types of electronic resonances, while there is little evidence for the efficacy of CAPs at treating states other than temporary anions. 
The CAP-augmented Hamiltonian has a purely discrete spectrum, whereas the continuous spectrum of the Hamiltonian is completely preserved, albeit rotated into the complex-plane, with complex scaling and the CBF method can be viewed as a finite basis representation of exterior complex scaling. From a more technical perspective, CBF methods tend to be more black-box. Different functional forms of the CAP are possible; box CAPs and CAPs based on Voronoi isosurfaces\cite{Sommerfeld:Voronoi:2015} are equally valid but yield different results. Similarly, while both methods require optimisation of some non-trivial parameter, there is no upper-limit on the CAP strength, $\eta$, whereas the optimal complex-scaling angle, $\theta$ is restricted to $0 < \theta < \pi/4$. Finally, the minimum in d$E/$d$\theta$ that defines the optimal scaling angle tends to be much more pronounced\cite{Utku:RICC2:2026} than the minimum in $\eta $d$E/$d$\eta$ that is sought in CAP methods.\cite{Gyamfi:CAP:2024,Sanov:CAP:2026} Likewise, optimal scaling angles are less sensitive to nuclear position than optimal CAP strengths, and real-space CAP onsets need to be reoptimised across potential energy surfaces.

In this communication, we present for the first time the theory of analytic nuclear gradients of the complex-energy CC and EOM-CC methods in a complex-scaled basis set as well as an efficient implementation in the singles-and-doubles (SD) approximation in the Q-Chem software.\cite{QChem}

Analytic derivatives have been recognised as the superior method for geometry optimisations since the pioneering work of Pulay demonstrated that not only are they more accurate than finite difference gradients, they are also computationally cheaper\cite{Pulay:HFGradient:1969}. By writing the gradient as a derivative of an appropriate Lagrangian functional, efficient gradients have been derived and implemented for the coupled-cluster (CC)\cite{Scheiner:CCSDGradient:1987,Gauss:CCGradient:1991} and equation-of-motion coupled-cluster (EOM-CC)\cite{Stanton:EOMCCGradient:1993,Stanton:EOMIPGradient:1994} methods for bound states making determination of highly accurate equilibrium structures and harmonic force constants of polyatomic molecules in their ground and excited states routine. For electronic resonances, there has been much less work, although generalisable analytic gradient expressions for CAPs have been introduced \cite{Benda:Gradient:2017} and recently extended to projected CAPs\cite{Mondal:ProjectedCAPGradient:2025}. The CAP gradient has been applied to geometry optimisations,\cite{Benda:Optimization:2018} minimum energy crossing point and exceptional point searches,\cite{Benda:MECPs:2018,Benda:EPs:2018} \emph{ab initio} molecular dynamics\cite{Gyamfi:AIMD:2022} and non-adiabatic couplings.\cite{Chatterjee:NACs:2023}
%There has been no work to date on complex energy gradients with CBFs.
 %We apply the new method to the geometry optimisations of a number of temporary anion resonances and core-ionised states. Likewise, we use numerical differentiation of the analytic gradient to compute for the first time harmonic frequencies for these states.

In the method of CBFs, a selection of Gaussian type orbitals have exponents scaled by the complex-coordinate prefactor, $e^{-2i\theta}$,
\begin{equation}
    \chi_\mu(\mathbf{r, A})=N_\mu(\theta)S_\mu(\mathbf{r-A})\cdot\exp\left[-\alpha e^{-2i\theta}(\mathbf{r}-\mathbf{A})^2\right].
\end{equation}
When every basis function is scaled, the method produces energies numerically identical to complex scaling of the Hamiltonian in that basis set. When the most diffuse basis functions of the basis set are scaled, the method is a finite basis representation of exterior complex scaling. In non-Hermitian quantum mechanics, the matrix elements of the Hamiltonian and all other operators, $\hat{O}$, are constructed using the c-product,\cite{Moiseyev:ResonanceProperties:1978,Moiseyev:Book:11} 
\begin{equation}\label{equation:cproduct}
    \langle\psi | \hat{O} | \phi \rangle_c = \int \psi(\mathbf{r})\hat{O}\phi(\mathbf{r})\mathrm{d}\mathbf{r},
\end{equation}
where the bra is \emph{not} complex-conjugated, unlike the standard Hermitian product metric. This leads to a Hamiltonian with a complex spectrum; for resonances, the observable quantities derive from the complex excitation energy $\Delta E = E_R - i\varGamma/2 = E_\mathrm{total} - E_\mathrm{threshold} - i\varGamma/2$. The real part of this energy, $E_R$, is the position of the resonance relative to the onset of the continuum and the width, $\varGamma$, is related to the decay lifetime by the energy-time uncertainty relation. 

Resonance wavefunctions are made square integrable using CBFs, provided that the scaling angle, $\theta$, is larger than the arctangent of the complex-valued momentum of the resonance $\theta > \arctan(\sqrt{2 \Delta E})$. In a complete basis, above this critical angle, the complex energy is independent of $\theta$. This is not, however, true in a finite basis, so in practical electronic structure calculations, an optimal scaling angle is determined by minimising the derivative of $\Delta E$ with respect to $\theta$,
\begin{equation}\label{equation:theta_opt}
    \theta_\mathrm{opt} = \argmin_\theta \sqrt{\left(\frac{\mathrm{d}E_{R}}{\mathrm{d}\theta}\right)^2+\left(\frac{1}{2}\frac{\mathrm{d\varGamma}}{\mathrm{d}\theta}\right)^2}.
\end{equation}

In CBF-EOM-CC, the resonance wavefunction is parametrised as $| \Psi \rangle = \hat{R}e^{\hat{T}} |\Psi_\mathrm{HF}\rangle$, where $|\Psi_\mathrm{HF}\rangle$ is the CBF Hartree-Fock ground state reference wavefunction, $\hat{T}$ is the
cluster operator and $\hat{R}$ is a linear operator which creates the target electronic states. $\hat{T}$ and $\hat{R}$ are defined analogously to standard EOM-CC theory but are in general complex-valued and constructed according to the c-product, Eq. \ref{equation:cproduct}.

As in the case of normal EOM-CC gradients, the CBF-EOM-CC gradient is most efficiently written as the derivative of an energy Lagrangian with respect to nuclear coordinates. In fact, the CBF-EOM-CC Lagrangian has exactly the same form as the normal EOM-CC Lagrangian except for the fact that (i) the effective one-electron and two-electron density matrices and integrals are complex-valued and (ii) the c-product is used:
\begin{equation}
\begin{aligned}
    \mathcal{L} &= \sum_{\mu\nu}D_{\mu\nu}^\mathrm{HF}h_{\mu\nu} + \frac{1}{2}\sum_{\mu\nu\sigma\rho}D_{\mu\nu}^\mathrm{HF}D_{\sigma\rho}^\mathrm{HF}\langle\mu\sigma\mid\mid\nu\rho\rangle_c + \sum_{\mu\nu}I_{\mu\nu}^\mathrm{HF}S_\mathrm{\mu\nu} \\
    &+ \sum_{\mu\nu}D_{\mu\nu}^\mathrm{CC}f_{\mu\nu}+ \sum_{\mu\nu\sigma\rho}\Gamma_{\mu\sigma\nu\rho}^\mathrm{CC}\langle\mu\sigma\mid\mid\nu\rho\rangle_c+ \sum_{\mu\nu}I_{\mu\nu}^\mathrm{CC}S_{\mu\nu}+V_\mathrm{nuc}.
\end{aligned}
\end{equation}
Here, $h_{\mu\nu}$, $f_{\mu\nu}$ and $S_{\mu\nu}$ are the elements of the one-electron Hamiltonian, Fock matrix and overlap integrals respectively, $\langle\mu\sigma\mid\mid\nu\rho\rangle_c$ are the antisymmetrised two-electron integrals, and $D$, $\Gamma$ and $I$ are the effective one-electron, two-electron and generalised energy-weighted density matrices respectively.\cite{gamma:vargamma} The nuclear repulsion energy $V_\mathrm{nuc}$ is the only real-valued quantity. 

%Naturally, the Lagrangian depends on the complex-scaling angle $\theta$, but we keep $\theta$ fixed while optimising the geometry, and then reoptimise the resonance energy as necessary. 

By making $\mathcal{L}$ stationary with respect to the molecular orbital coefficients and CC amplitudes,\cite{Helgaker:Gradient:1988,Jorgensen:MP2Gradient:1988} we arrive at the following expression for the complex-valued gradient:
\begin{equation}\label{equation:gradient}
    \begin{aligned}
        \frac{\mathrm{d}E}{\mathrm{d}R_k} &= \frac{\mathrm{d}E_\mathrm{total}}{\mathrm{d}R_k} -\frac{i}{2} \frac{\mathrm{d}\varGamma}{\mathrm{d}R_k}
        \\ &= \sum_{\mu\nu}\left[D_{\mu\nu}^\mathrm{HF} + D_{\mu\nu}^\mathrm{CC} \right] \frac{\mathrm{d} h_{\mu\nu}}{\mathrm{d} R_k} 
        \\ &+ \sum_{\mu\nu\sigma\rho}\left[\frac{1}{2}D_{\mu\nu}^\mathrm{HF}D_{\sigma\rho}^\mathrm{HF} + D_{\mu\nu}^\mathrm{CC}D_{\sigma\rho}^\mathrm{HF} + \Gamma_{\mu\sigma\nu\rho}^\mathrm{CC}\right]\frac{\mathrm{d}\left(\mu\sigma\mid\mid\nu\rho\right)}{\mathrm{d} R_k} 
        \\ &+ \sum_{\mu\nu}\left[I_{\mu\nu}^\mathrm{HF} + I_{\mu\nu}^\mathrm{CC}\right]\frac{\mathrm{d} S_\mathrm{\mu\nu}}{\mathrm{d} R_k} + \frac{\mathrm{d} V_\mathrm{nuc}}{\mathrm{d} R_k}.
    \end{aligned}
\end{equation}
The real part of the gradient is responsible for the forces that propagate the nuclei on the complex-valued potential energy surface, and the imaginary part governs change of the electronic decay rate.\cite{Moiseyev:Forces:2017} We emphasise here that the quantity relevant to nuclear dynamics and geometry optimisations \emph{etc.} is the \emph{total} electronic energy and not the resonance position.

Of the various derivative integrals in Eq. \ref{equation:gradient}, the electron repulsion integrals (ERIs) are the most difficult to handle efficiently. The starting point for all ERI derivatives is computation of the Boys function,\cite{Boys:BoysFunction:1950} which for any ERI with at least one CBF is complex-valued. In our implementation, we use the algorithm proposed by White \emph{et al.}\cite{White:CBF:2015} for undifferentiated ERIs where we note that the Boys function is computed on the fly rather than using the interpolation schemes generally used for the real-valued Boys function. Then, the target ERI derivatives can be formed using all of the standard recurrence relations,\cite{McMurchie:TEIs:1978,Obara:TEIs:1986,HeadGordon:TEIs:1988} which are valid for arbitrary derivative integrals over Gaussian type basis functions with complex exponents provided that the real part of the exponent is negative. In the context of CBFs, this condition is equivalent to the requirement that $\theta < \pi/4$. 
%We form the ERI derivatives using the Head-Gordon-Pople algorithm for integral derivatives,\cite{HeadGordon:TEIs:1988} modified to account for the use of complex intermediates in the recurrence relations. 
We have implemented recurrence relations for the first and second derivatives of Hamiltonian matrix elements with complex-scaled exponents in Q-Chem's \texttt{libqints} library with density contractions handled in the \texttt{libposthf} and \texttt{libfock} libraries. Intermediates are computed with the Head-Gordon-Pople algorithm\cite{HeadGordon:TEIs:1988} for the electron-repulsion and electron-nucleus-attraction integral derivatives, the Gao-Thorvaldsen-Ruud algorithm\cite{Gao:Integrals:2011} for the overlap integral derivatives, and the Taketa-Huzinaga-Ohata algorithm\cite{Taketa:Integrals:1966,Petersson:Integrals:2009} for the kinetic energy integral derivatives. The usual permutation symmetries of the derivative integrals remain valid in the complex case due to the c-product metric, Eq. \ref{equation:cproduct}.

In our implementation, the complex effective CC and EOM-CC density matrices are computed first in the molecular orbital basis, back-transformed into the atomic-orbital basis and then contracted with the one- and two-electron integral derivatives directly without the need for storage of any four-index atomic-orbital quantities in the random-access memory, as is done in all mature implementations of real-valued EOM-CC gradients. In the ionisation potential (IP)\cite{Stanton:EOMIPGradient:1994} and electron attachment (EA)\cite{Nooijen:EOMEA:1995} variants of EOM-CCSD, computation of the $\hat{R}$ amplitudes scales only with the fifth power of the basis set size. Also, formation of the two-electron integral derivatives and contraction of these with the two-particle density matrix scales only quartically with basis set size. This means that the rate-determining steps in the evaluation of EOM-IP-CCSD and EOM-EA-CCSD gradients are the $\mathcal{O}(o^2v^4)$ computations of the CC amplitudes and response Lagrange multipliers and the two-particle density matrix. Given that for CBFs and CAPs, these are computed using complex algebra in the molecular orbital basis, CBF-EOM-CCSD gradients are not more expensive than CAP-EOM-CCSD gradients, although they are more expensive than standard EOM-CCSD gradients using real wavefunctions. The density matrices are computed in Q-Chem's \texttt{ccman2} library. We reuse, from CAP-EOM-CCSD gradients, all electron complex energy orbital response Lagrange multipliers\cite{Benda:Gradient:2017} and have implemented complex energy orbital response for frozen-core gradients. The underlying tensor algebra is performed with \texttt{libtensor} which exploits spin, permutation, and point group symmetry.\cite{Epifanovsky:libtensor:2013}

Given that the basis functions depend explicitly on the scaling angle $\theta$, the Lagrangian and derivative expressions do so implicitly. Thus, one might argue that a geometry optimisation of a resonance state with a CBF method might either require that $\mathcal{L}$ be made stationary with respect to $\theta$ or include the derivatives of the Hamiltonian with respect to $\theta$ in the gradient expression in order to introduce $\theta$ as an additional optimisation parameter. %However, both approaches are problematic due to the criterion in Eq. \ref{equation:theta_opt}: the \emph{complex excitation} energy is made stationary with respect to $\theta$ but the \emph{real total} energy is minimised with respect to nuclear positions. Thus to achieve $(a)$, requires a Lagrangian stationary in both the total and excitation energy. Alternatively, achieving $(b)$ would require iterative reoptimisation of the structure, then $\theta_\mathrm{opt}$, then the structure again and so on. 
However, given that $\theta$ is a one-dimensional parameter, it is much easier to optimise this parameter ``by hand'' as in CBF single-point calculations. Thus in our protocol, we optimise the geometry at a given $\theta$ and then determine $\theta_\mathrm{opt}$ from a number of single point calculations at the new geometry. Then, if $\theta_\mathrm{opt}$ has changed, the geometry is reoptimised at the new fixed $\theta$. This is similar to the procedure proposed for CAP geometry optimisations, where $\eta_\mathrm{opt}$ and the CAP onset both need to be redetermined every time the geometry is re-optimised.\cite{Benda:Gradient:2017,Benda:Optimization:2018} In fact, our numerical results show that neither for temporary anions nor core-ionised states did $\theta_\mathrm{opt}$ ever change by more than a few degrees and in some cases did not change at all (within steps of 1$^\circ$). Therefore, only one or two cycles of this process are needed to converge a structure which is rather user friendly. 

We demonstrate the use of CBF-EOM-CC analytic nuclear gradients by computing equilibrium structures and vibrational frequencies for a number of temporary anion resonances and core-ionised states of small molecules. For the anions, we computed the structures using EOM-EA-CCSD with both CAPs and CBFs for comparison. For the cations, we computed the structures with EOM-IP-CCSD with CBFs and also with the core-valence separation (CVS)\cite{Cederbaum:CVS:1980,Coriani:CVS:2015,Frati:CVSGradient:2019} approach as implemented in the CFOUR software\cite{CFOUR} where EOM-CC amplitudes describing Auger decay are zeroed out. %For technical reasons, we used energy-only CVS gradient calculations. 
For computing structures and frequencies of the parent neutral molecules, we use standard CCSD with real basis functions. %Additionally in this work, we have implemented Dyson orbitals within the CBF-EOM-CCSD in order to visualise the resonance wavefunctions.

The choice of an appropriately large basis set is especially important in computations on electronic resonances. In the context of CBFs, one needs an adequate real basis set which can describe the molecular interaction region, as well as adequate complex basis functions which approximate the asymptotic form of the exterior complex scaled wavefunction. For temporary anion resonances, we use the aug-cc-pVTZ basis\cite{Dunning:pVTZ:1989,Kendall:aug:1992} augmented by three additional shells of $s, p$ and $d$ functions on all non-hydrogen atoms. The exponents are generated in an even-tempered manner by recursively dividing the exponents of the most diffuse functions by 2. In the CBF calculations, the first of these shells is unscaled, and the latter two are scaled. We performed both frozen-core and all electron calculations on the anions.

In calculations on core-ionised states, a high cardinality of the basis is essential to capture the coupling to the continuum: since these states are Feshbach resonances, the valence electrons directly participate in the decay. Moreover, relatively steep complex functions are required in order to account for the fast outgoing electron. In these calculations, we used the cc-pCVTZ(5sp)+3$spd$ basis\cite{Dunning:pVTZ:1989,Woon:PolarizedCore:1995} where the parentheses indicate that real-valued $s$ and $p$ shells from the cc-pCV5Z basis and the $d$ and $f$ functions from cc-pCVTZ are used. We use custom complex  exponents on all non-hydrogen atoms optimised for Auger decay\cite{Matz:Auger:2022}; they are given in the SI. Since we want to capture the relaxation of the core due to the ionisation, all electrons are correlated in these calculations.

To calculate the vibrational frequencies, we use numerical differentiation of the real part of the analytic gradient. For use with the Q-Chem software, we have developed a series of small standalone codes which allow the exploitation of point group symmetry in these calculations. As has been discussed before and implemented elsewhere,\cite{CFOUR} there are numerous ways in which this can significantly speed up frequency calculations.\cite{Stanton:FrequencyReview:2000} Our code generates symmetry adapted linear combinations of the mass-weighted Cartesian displacement vectors for a given molecule within the largest Abelian subgroup of the point group. Rotations and translations are projected out to create a set of 3N$_\mathrm{atom}-$6 displacement vectors that belong to irreducible representations of the point group. For totally symmetric displacements, each vibrational coordinate requires two force calculations to compute a central difference derivative. For nontotally symmetric displacements, we exploit the fact that the change in the gradient in both directions of the displacement is identical. %The use of the stand alone codes mean that frequency calculations can be considerably sped up in three main ways: (i) the force calculations can be performed with Abelian point group symmetry using Q-Chem's \texttt{libtensor} library, (ii) the 6 rotational and translational modes are not included in the set of displacements unlike in codes which simply displace the atoms in Cartesian space and (iii) the use of separate input files for force calculations means that the whole procedure can be trivially parallelised across displacements on a multinode high performance computing cluster. 
We note that the operation count for computing the CCSD Hessian scales linearly with the number of atoms, both for fully analytical differentiation as well as for numerical differentiation of analytical gradients.\cite{Gauss:Properties:2000} Hence, the approach involving separate force calculations is faster than analytic Hessians since the whole procedure can be trivially parallelised across displacements on a multinode high performance computing cluster. In CBF frequency calculations, we fix $\theta$ at the optimal value for the resonance equilibrium structure for simplicity and in all frequency calculations, we use a step size of 0.001 {\AA} for numerical differentiation.

 \begin{table*}
     \caption{Bond lengths, $R$ ({\AA}), bond angles, $\angle$ ($^\circ$), dihedral angles, $\tau$ ($^\circ$), and zero-point vibrational energies, ZPVE (eV) of the various temporary anions and their parent neutral molecules. Computed with EOM-EA-CCSD, all electron (AE) and frozen-core (FC). For HCOOH, $O_a$ and $O_b$ refer to the carbonyl and hydroxyl oxygen nuclei respectively. Structures optimized with all electron full CAP taken from Ref \citenum{Benda:Optimization:2018}. Structures optimised with projected CAP taken from Ref \citenum{Mondal:ProjectedCAPGradient:2025}.}
     \label{tab:anion_structures}
     \centering
     \begin{ruledtabular}
    \begin{tabular}{lcccccccc}
         CH$_2$O & {} & {} & $R$(CO) & $R$(CH) & $\angle$(HCO) & $\angle$(HCH) & $\tau$(HCHO) & ZPVE \\ \hline
         Neutral & AE & $C_{2v}$ & 1.200 & 1.097 & 121.8 & 116.5 & 180.0 & 0.737 \\
         {} & FC & $C_{2v}$ & 1.205 & 1.101 & 121.7 & 116.5 & 180.0 & 0.734 \\ \hline
         CBF & AE  & $C_s$ & 1.286 & 1.101 & 122.3 & 114.6 & 170.4 & 0.662 \\
         {} & FC  & $C_s$ & 1.289 & 1.105 & 122.3 & 114.5 & 169.6 & 0.662 \\ %\hline
         Full CAP & FC  & $C_{2v}$ & 1.273 & 1.103 & 122.4 & 115.2 & 180.0 & $\cdots$ \\
         Full CAP/aug-cc-pVTZ+3p3d & AE & $C_{2v}$ & 1.275 & 1.098 & 122.4 & 115.2 & 180.0 & $\cdots$ \\
        % \multicolumn{2}{l}{Proj. CAP/SA-15-CASSCF/cc-pVTZ+X[4p]}& $C_s$ & 1.278 & 1.110 & 120.3 & 112.7 & 151.8 & $\cdots$\\
         \multicolumn{2}{l}{Proj. CAP-SA-19-CASSCF}%/cc-pVTZ+X[2s5p2d]} 
         & $C_s$ & 1.268 & 1.098 & 121.0 & 113.1 & 156.1 & $\cdots$\\
         \multicolumn{2}{l}{Proj. CAP-MRCIS/SA-7-CASSCF}%/cc-pVDZ+[3p]} 
         & $C_{2v}$ & 1.283 & 1.102 & 121.9 & 116.2 & 180.0 & $\cdots$\\
     \end{tabular}
     \begin{tabular}{lcccccccc}
          C$_2$H$_4$ & {} & {} & $R$(CC) & $R$(CH) & $\angle$(HCC) & $\angle$(HCH) & $\tau$(HCCH) & ZPVE \\ \hline
        Neutral & AE & $D_{2h}$ & 1.328 & 1.078 & 117.1 & 121.4 & 0 & 1.395 \\ 
        {} & FC & $D_{2h}$ & 1.332 & 1.082 & 117.0 & 121.5 & 0 & 1.392 \\ \hline
         CBF & AE & $C_{2h}$ & 1.418 & 1.084 & 120.6 & 115.1 & 23.0 & 1.244 \\
         {} & FC & $C_{2h}$ & 1.423 & 1.089 & 120.6 & 114.9 & 23.5 & 1.241 \\ %\hline
         Full CAP & FC & $C_{2h}$ & 1.404 & 1.085 & 121.9 & 116.2 & 0.3 & $\cdots$ \\ 
         Full CAP/aug-cc-pVTZ+3p & AE & $C_{2h}$ & 1.414 & 1.084 & 120.7 & 115.4 & 21.0 & $\cdots$ \\
         \multicolumn{2}{l}{Proj. CAP-SA-15-CASSCF} & $C_{2h}$ & 1.429 & 1.080 & 118.2 & 113.2 & 37.2 & $\cdots$  \\
         \multicolumn{2}{l}{Proj. CAP-MRCIS/SA-6-CASSCF} & $C_{2h}$ & 1.426 & 1.094 & 119.4 & 114.2 & 30.9 & $\cdots$
     \end{tabular}
          \begin{tabular}{lcccccccccccc}
          HCOOH & {} & {} & $R$(CO$_a$) & $R$(CO$_b$) & $R$(CH) & $R$(OH) & $\angle$(OCO) & $\angle$(O$_a$CH) & $\angle$(O$_b$CH) & $\angle$(COH) & $\tau$(HOCH) & ZPVE \\ \hline
        Neutral & AE & $C_s$ & 1.193 & 1.336 & 1.089 & 0.963 & 125.1 & 124.9 & 109.9 & 107.2 & 180.0 & 0.940 \\
        {} & FC & $C_s$ & 1.198 & 1.341 & 1.093 & 0.966 & 125.0 & 125.0 & 110.1 & 107.0 & 180.0 & 0.934\\ \hline
         CBF & AE & $C_1$ & 1.265 & 1.443 & 1.114 & 0.964 & 115.5 & 120.8 & 107.8 & 100.5 & 135.9 & 0.804 \\
         {} & FC & $C_1$ & 1.271 & 1.451 & 1.120 & 0.967 & 115.4 & 120.6 & 107.6 & 100.4 & 135.0 & 0.801\\ 
         Full CAP/aug-cc-pVTZ+3p3d & AE & $C_{1}$ & 1.263 & 1.433 & 1.106 & 0.965 & 116.7 & 121.0 & 108.2 & 101.4 & 136.9 & $\cdots$ \\
         \multicolumn{2}{l}{Proj. CAP-SA-16-CASSCF}%/cc-pVTZ+X[2s5p2d]} 
         & $C_1$ & 1.265 & 1.430 & 1.123 & 0.964 & 113.8 & 119.0 & 108.9 & 103.5 & 115.5 & $\cdots$ \\
         \multicolumn{2}{l}{Proj. CAP-MRCIS/SA-8-CASSCF}%/cc-pVTZ+X[2s5p2d]} 
         & $C_1$ & 1.266 & 1.465 & 1.143 & 0.949 & 114.8 & 119.1 & 104.9 & 101.5 & 136.6 & $\cdots$
     \end{tabular}
     \end{ruledtabular}
\end{table*}

Table \ref{tab:anion_structures} details the structural parameters and zero-point vibrational energies (ZPVE) of the temporary anions of CH$_2$O, HCOOH and C$_2$H$_4$. Individual harmonic frequencies with assignments are reported in the SI. All three anions are $\pi^*$-like at the parent neutral structure and hence the effective bond-order of the relevant double bond is formally decreased. All anions are subject to symmetry breaking out-of-plane distortions with respect to the parent neutral molecule. For the formaldehyde and ethylene anions, the bending is moderate and the most prominent structural change is a significant stretch of the double bonds by \textit{ca.} 0.09 {\AA} in both cases. For the formic acid anion, the twisting of the hydroxy group out of the plane is very pronounced and consequently, the C-OH bond is also stretched by \textit{ca.} 0.11 \AA. This can be rationalised by a weaker mesomeric effect since the hydroxy group is no longer in resonance with the $\pi$-system.
We note that since the Siegert energy of an electronic resonance possesses an imaginary part, there is always a finite energy gap between the electronic resonance and the continuum of elastic scattering states. This means that derivative coupling between the resonance and the continuum is non-singular even in a complete basis and hence, symmetry lowering distortions are driven by a non-adiabatic coupling force.\cite{Moorby:Swave:2024,Moorby:Swave:2026}
Bond lengths and bond angles are in close agreement between CBFs and CAPs. Both full and projected CAP methods also predict symmetry lowering for the ethylene and formic acid anions with respect to the parent neutral molecules. However, for the formaldehyde anion, CAP-EOM-EA-CCSD predicts a planar geometry in agreement with previous work.\cite{Benda:Gradient:2017,Benda:Optimization:2018} We note though that in Ref \citenum{Mondal:ProjectedCAPGradient:2025}, the authors reported a non-planar structure for CH$_2$O$^-$ with projected-CAP-CASSCF/cc-pVTZ+X[2s5p2d] and a planar one with projected-CAP-MRCIS/cc-pVDZ+[3p].

In all three molecules, the zero-point vibrational energy is lower for the resonance than the parent neutral molecule; given that the excess electron occupies an anti-bonding orbital this is unsurprising. To be sure that complex scaling the basis functions does not distort the shape of the potential energy surfaces, we recomputed the equilibrium structure and harmonic frequencies of neutral CH$_2$O with CBF-CCSD at a complex scaling angle of 26$^\circ$ \emph{i.e.} the optimal angle for the resonance at its equilibrium structure. The equilibrium structure changes negligibly and changes in all vibrational frequencies are less than 10 cm$^{-1}$.

\begin{table}
    \caption{Theoretical and experimental vertical electron affinities (VEA), vertical detachment energies (VDE), adiabatic electron affinities (AEA), and decay widths at the neutral (0) and anion ($\mathrm{res}$) structures. All values in eV. Adiabatic affinities include zero-point vibrational corrections}
    \label{tab:anion_energies}
    \centering
    \begin{ruledtabular}
    \begin{tabular}{cccccccc}
        {} & {} & {} & VEA & VDE & AEA & $\varGamma_0$ & $\varGamma_\mathrm{res}$ \\ \hline
        \multirow{3}{*}{CH$_2$O$^-$} & CBF & AE & $-$1.197 & $-$0.648 & $-$0.875 & 0.579 & 0.296 \\
        {} & {} & FC & $-$1.180 & $-$0.634 & $-$0.856 & 0.565 & 0.287 \\ 
%        {} & CAP & FC & $-$1.190 & --- & --- & 0.374 & --- \\ 
        {} & Expt. & &  $-$0.86\cite{Burrow:CH2OAnion:1976} & $\cdots$ & $-$0.65\cite{Burrow:CH2OAnion:1976} & \multicolumn{2}{c}{$0.2$--$0.4$\cite{Burrow:CH2OAnion:1976}} \\
        \hline
        \multirow{3}{*}{C$_2$H$_4^-$} & CBF & AE & $-$2.106 & $-$1.539 & $-$1.746 & 0.788 & 0.413 \\
        {} & {} & FC & $-$2.088 & $-$1.516 & $-$1.725 & 0.777 & 0.402 \\ 
%        {} & CAP & FC & --- & --- & --- & --- & --- \\
        {} & Expt. & {} & $-1.76$\cite{Sanche:C2H4Anion:1973} & $\cdots$ & $-$1.55\cite{Burrow:C2H4Anion:1975} & \multicolumn{2}{c}{$0.3$--$0.7$\cite{Sanche:C2H4Anion:1973}} \\
        \hline
        \multirow{3}{*}{HCOOH$^-$} & CBF & AE & $-$2.325 & $-$0.332 & $-$1.386 & 0.521 & 0.123 \\
        {} & {} & FC & $-$2.285 & $-$0.260 & $-$1.335 & 0.509 & 0.109 \\ 
        %{} & CAP & FC & --- & --- & --- & --- & --- \\ 
        {} & Expt. & {} &  $-1.73$\cite{Aflatooni:HCOOHAnion:2001} & $\cdots$ & $-1.3$\cite{Tronc:HCOOH:1987} & \multicolumn{2}{c}{$\cdots$} \\ 
    \end{tabular}
    \end{ruledtabular}
\end{table}

For all anions, structural relaxation leads to a significant decrease in the decay width, as Table \ref{tab:anion_energies} shows. At their equilibrium structures, all of the anions remain unbound, albeit closer to the detachment threshold.

From a technical perspective, we found the optimisations using CBFs to be more user friendly than with CAPs. Beyond the need to reoptimise only one parameter in the CBF case, the CBF gradient is rotationally and translationally invariant. This is not true of the CAP gradient due to the cuboidal box.\cite{Benda:Optimization:2018} This often meant that in CAP optimisations, the gradient could not be back-transformed into internal coordinates, and the optimiser was forced to proceed in Cartesian coordinates. As a result, the optimisation converged quite slowly with CAPs and the maximum in the gradient often levelled off, whereas for CBFs, it took no user intervention to converge the maximum in the gradient below $10^{-8}$ Eh bohr$^{-1}$. For all states, the effect of the frozen core on the equilibrium structure is negligible and the same order of magnitude as for bound state optimisations;  zero-point energies vary by $ca.$ 5 meV only. For HCOOH$^-$, one geometry cycle took around 44 CPU hours with frozen core and 55 hours without.

 \begin{table*}
     \caption{Bond lengths, $R$ ({\AA}), bond angles, $\angle$ ($^\circ$), and zero-point vibrational energies, ZPVE (eV) of the various core-ionised states. For the b$_{3u}^{-1}$ state of C$_2$H$_4$, the core hole localises onto the atom labelled C$_b$.}
     \label{tab:auger_structures}
     \centering
     \begin{ruledtabular}
         \begin{tabular}{lccccc}
         H$_2$O & {} & {} & $R$(OH) &$\angle$(HOH) & ZPVE \\ \hline
         Neutral & Real & $C_{2v}$ & 0.9549 & 104.30 & 0.593 \\
         1a$_1^{-1}$ & CBF & $C_{2v}$ & 0.9459 & 116.53 & 0.570  \\ 
         {} & CVS & $C_{2v}$ & 0.9459 & 116.56 & $\cdots$ \\ 
     \end{tabular}
    \begin{tabular}{lcccccc}
         CH$_2$O & {} & {} & $R$(CO) & $R$(CH) & $\angle$(HCO) & ZPVE \\ \hline
         Neutral & Real & $C_{2v}$ & 1.1985 & 1.0986 & 121.80 & 0.739\\
         1a$_1^{-1}$ & CBF & $C_{2v}$ & 1.2059 & 1.0805 & 117.28 & 0.782 \\ 
        {} & CVS & $C_{2v}$ & 1.2081 & 1.0803 & 117.24 & $\cdots$ \\
         2a$_1^{-1}$ & CBF & $C_{2v}$ & 1.1764 & 1.0436 & 119.95 & 0.818\\ 
         {} & CVS & $C_{2v}$ & 1.1763 & 1.0433 & 119.95 & $\cdots$
     \end{tabular}
     \begin{tabular}{lccccccccc}
          C$_2$H$_4$ & {} & {} & $R$(CC) & $R$(C$_a$H) & $R$(C$_b$H) & $\angle$(HC$_a$C$_b$) & $\angle$(HC$_b$C$_a$) & ZPVE \\ \hline
        Neutral & Real & $D_{2h}$ & 1.3264 & 1.0792 & 1.0792 & 121.51 & 121.51 & 1.399 \\
         1a$_\mathrm{g}^{-1}$ & CBF & $D_{2h}$ & 1.2839 & 1.0504 & 1.0504 & 120.52 & 120.52 & 1.738 \\ 
         %{} & Real & $D_{2h}$ & --- & --- & --- & --- & --- & --- \\
         {} & CVS & $D_{2h}$ & 1.2835 & 1.0502 & 1.0502 & 120.50 & 120.50 & $\cdots$ \\
         1b$_{3\mathrm{u}}^{-1}$ & CBF & $C_{2v}$ & 1.2806 & 1.0747 & 1.0275 & 119.89 & 121.23 & 1.506\\
         {} & CVS & $C_{2v}$ & 1.2807 & 1.0747 & 1.0272 & 119.87 & 121.22 & $\cdots$
         %{} & Real & $C_{2v}$ & 1.281 & 1.069 & 1.041 & 120.2 & 120.9 & 1.410
     \end{tabular}
     \end{ruledtabular}
\end{table*}

Table \ref{tab:auger_structures} details the structural parameters and ZPVEs of the K-shell ionised states of water, formaldehyde and ethylene. The structures computed with CBFs agree remarkably well with those computed using core-valence separation, validating the CBF results. For water, the zero-point energy decreases on core-ionisation due to a decrease of the frequency associated with HOH bending. Since electron density from the oxygen lone pairs partially fills the core hole, the HOH bond angle is less stiff, and thus the zero-point energy lowers. For the formaldehyde and ethylene molecules, the effect of core-ionisation is a decrease of bond lengths and substantial increase of vibrational frequencies. This can be rationalised by the equivalent core approximation:\cite{Shirley:EquivalentCore:1972} a bond between a hydrogen atom and 1s ionised carbon behaves like a nitrogen-hydrogen bond. In C$_2$H$_4$, the near degeneracy of the two core molecular orbitals results in a strong pseudo-Jahn-Teller effect that distorts the 1b$_\mathrm{3u}^{-1}$ state. The resultant $C_{2v}$ structure has a localised core-hole on the carbon atom labelled $C_b$ in Table \ref{tab:auger_structures}. This creates a very steep harmonic potential along this normal coordinate in the totally symmetric state; it acquires one single frequency of \emph{ca.} 7000 cm$^{-1}$along the b$_\mathrm{3u}$ normal coordinate associated with the symmetry breaking. This leads to an increase in the zero-point energy for the 1a$_\mathrm{g}^{-1}$ state by \emph{ca.} 0.2 eV.
%Clearly, a theory of non-adiabatic coupling in the continuum is needed in the future.

\begin{table}
    \caption{Theoretical and experimental vertical ionisation energies (IE, eV) at the neutral (0) and resonance ($\mathrm{res}$) equilibrium structures, adiabatic ionisation energy (AIE, eV) and decay widths (meV). Adiabatic ionisation energies include zero-point vibrational corrections}
    \label{tab:auger_energies}
    \centering
    \begin{ruledtabular}
    \begin{tabular}{cccccccc}
        {} & {} & {} & IE$_0$ & IE$_\mathrm{res}$ & AIE & $\varGamma_0$ & $\varGamma_\mathrm{res}$ \\ \hline
                \multirow{2}{*}{H$_2$O} & \multirow{2}{*}
                {1a$_1^{-1}$} & CBF & 541.360 & 541.189 & 541.250 & 180.2 & 170.1 \\ 
        {} & {} & Expt. & 539.8\cite{Sankari:WaterAuger:2003} & $\cdots$ & $\cdots$ & \multicolumn{2}{c}{160$\pm 5$\cite{Sankari:WaterAuger:2003}} \\ \hline
        \multirow{4}{*}{CH$_2$O} & \multirow{2}{*}{1a$_1^{-1}$} & CBF & 541.330 & 541.192 & 541.290 & 183.2 & 171.0 \\ 
        {} & {} & Expt. & 539.48\cite{Kuramoto:CH2OAuger:2005} & $\cdots$ & $\cdots$ & \multicolumn{2}{c}{$\cdots$}\\
        {} & \multirow{2}{*}{2a$_1^{-1}$} & CBF & 295.871 & 295.571 & 295.787 & 79.9 & 84.1 \\
        {} & {} & Expt. & 294.47\cite{Kuramoto:CH2OAuger:2005} & $\cdots$ & $\cdots$ & \multicolumn{2}{c}{83$\pm 7$\cite{DeFanis:CH2OAuger:2005}} \\ \hline
        \multirow{3}{*}{C$_2$H$_4$} & 1a$_\mathrm{g}^{-1}$ & CBF & 292.160 & 291.934 & 292.417 & 78.2 & 76.8 \\ 
        {} & 1b$_{3\mathrm{u}}^{-1}$ & CBF & 292.210 & 291.788 & 292.090 & 73.9 & 83.0 \\
        {} & $\cdots$ & Expt. & 290.6\cite{Hitchcock:C2H4Auger:1977} & $\cdots$ & $\cdots$ & \multicolumn{2}{c}{85$\pm 10$\cite{Kempgens:C2H4Auger:1998}} 
    \end{tabular}
    \end{ruledtabular}
\end{table}

Table \ref{tab:auger_energies} details the vertical and adiabatic core-ionisation energies of the studied systems. Unlike in the anionic case, structural relaxation of the core-ionised state in some cases causes the decay width to increase somewhat. This is accompanied by significant contraction of CH bonds, indicating that the migration of the electron density from the CH bonding molecular orbitals towards the core hole leads to an enhancement of the Auger decay process.

We additionally attempted to perform optimisations for core-ionised states using real-valued EOM-IP-CCSD. %Here, the Davidson solver took $>$100 iterations, whereas with CBFs it took fewer than 30 iterations to converge the residual norm to machine precision. 
In both Q-Chem and CFOUR, at a number of geometry steps, the Davidson solver converged to pseudocontinuum states with significant double excitation character, $||\hat{R}_2||^2 > 0.4$. This in turn led to huge oscillations in the total energy between geometry iterations and no minima could be found, illustrating the unbound nature of core-ionised states. 

In conclusion, we have presented the general theory for the analytic nuclear gradient of complex-valued Siegert energies with EOM-CC and complex basis functions. For anions, CBF-EOM-CC optimisations are more black box than CAP-EOM-CC optimisations. For core-ionised states, geometry optimisations perfom similarly to the well-established CVS-EOM-IP-CCSD approach. One particular strength of the CBF approach is its applicability to various types of electronic decay processes. We note that inner-valence ionised states can also be subject to autoionisation,\cite{Molloy:LLLDecay:2006,Eland:LLLDecay:2008,Arvinder:LLLDecay:2008,Idebohn:LLLDecay:2026,Schepens:LLLDecay:2026} while the validity of the core-valence separation for such states is more questionable. At the same time, inner-valence ionisation is, on average, more likely than core-ionisation to lead to structural changes, making such states an interesting application of CBF-EOM-CC gradients.
                
An attractive extension of the work is for the evaluation of vibrational broadening in Auger spectra.\cite{Cederbaum:AugerBroadening:1991} In diatomics, the consideration of vibrational effects considerably improves agreement between \emph{ab initio} and experimental Auger spectra.\cite{Camps:RICCSD:2025} Also of interest is the implementation of analytic gradients for lower-cost electronic structure methods for example CBF-RICC2 which has recently been applied to temporary anions.\cite{Utku:RICC2:2026} 

\begin{acknowledgments}
T.-C.J. acknowledges funding from the European Research Council (ERC) under the European Union’s Horizon 2020 research and innovation program (Grant Agreement No. 851766) and the KU Leuven internal funds (Grant No. C14/22/083).
R.E.M. thanks Simen Camps and Dr. Paul Albrecht for valuable discussions, Yifan Jiang for help with residual solvers for the EOM equations, and Dr. Anthuan Ferino P{\' e}rez for help with CVS calculations in CFOUR.
\end{acknowledgments}

\section*{Data Availability Statement}
The data that support the findings of this study are available within the article and its supplementary material.

\section*{Supplementary Material}
Coordinates of all optimised structures in Z-matrix format, harmonic vibrational frequencies and their assignments, optimal complex-variable parameters and additional basis function exponents used in calculations.

\section*{Author Declarations}
The authors have no conflicts to disclose.

\bibliography{cbf_gradients}% Produces the bibliography via BibTeX.

@PREAMBLE{
 "\providecommand{\noopsort}[1]{}" 
 # "\providecommand{\singleletter}[1]{#1}%" 
}

@string{acs = "Acta Chem. Scand."}

@string{atoms = "Atoms"}

@string{arpc = "Annu. Rev. Phys. Chem."}

@string{cpl = "Chem. Phys. Lett."}

@string{cmp = "Commun. Math. Phys."}

@string{ijqc = "Int. J. Quant. Chem."}

@string{jcc = "J. Comput. Chem."}

@string{jcp = "J. Chem. Phys."}

@String{jctc = "J. Chem. Theory Comput."}

@string{jpB = "J. Phys. B: At. Mol. Opt. Phys."}

@string{jpb = "J. Phys. B"}

@string{jpca = "J. Phys. Chem. A"}

@string{jpcl = "J. Phys. Chem. Lett."}

@string{mp = "Mol. Phys."}

@string{nature = "Nature"}

@string{pccp = "Phys. Chem. Chem. Phys."}

@string{pra = "Phys. Rev. A"}

@string{prA = "Phys. Rev. A"}

@string{prl = "Phys. Rev. Lett."}

@string{prp = "Phys. Rep."}

@article{Simons:Anions:2011,
   author = "Simons, Jack",
   title = "Theoretical Study of Negative Molecular Ions", 
   journal= "Annu. Rev. Phys. Chem.",
   year = "2011",
   volume = "62",
   number = "Volume 62, 2011",
   pages = "107-128",
   doi = "https://doi.org/10.1146/annurev-physchem-032210-103547",
   publisher = "Annual Reviews",
   issn = "1545-1593",
   type = "Journal Article"
  }

@inbook{Herbert:Anions:2015,
author = {Herbert, John M.},
publisher = {John Wiley \& Sons, Ltd},
isbn = {9781118889886},
title = {The Quantum Chemistry of Loosely-Bound Electrons},
booktitle = {Reviews in Computational Chemistry Volume 28},
chapter = {8},
pages = {391-517},
doi = {https://doi.org/10.1002/9781118889886.ch8},
year = {2015}
}

@article{Alizadeh:Anions:2015,
   author = "Alizadeh, Elahe and Orlando, Thomas M. and Sanche, Léon",
   title = "Biomolecular Damage Induced by Ionizing Radiation: The Direct and Indirect Effects of Low-Energy Electrons on {DNA}", 
   journal= "Annu. Rev. Phys. Chem.",
   year = "2015",
   volume = "66",
   number = "Volume 66, 2015",
   pages = "379-398",
   doi = "https://doi.org/10.1146/annurev-physchem-040513-103605",
   publisher = "Annual Reviews",
   issn = "1545-1593",
   type = "Journal Article"
  }

@book{Agarwal:Xrays:2013,
  author    = {B. K. Agarwal},
  title     = {X-Ray Spectroscopy: An Introduction},
  publisher = {Springer},
  address   = {Berlin, Heidelberg},
  year      = {2013}
}

@article{Norman:Xrays:2018,
    author = {Norman, Patrick and Dreuw, Andreas},
    title = {Simulating {X}‑ray Spectroscopies and Calculating
Core-Excited States of Molecules},
    journal = {Chem. Rev.},
    volume = {118},
    number = {15},
    pages = {7208-7248},
    year = {2018},
    month = {06},
    issn = {0009-2665},
    doi = {10.1021/acs.chemrev.8b00156},
}

@article{Zimmermann:Xrays:2020,
title = {Modern {X}-ray spectroscopy: {XAS and XES} in the laboratory},
journal = {Coord. Chem. Rev.},
volume = {423},
pages = {213466},
year = {2020},
issn = {0010-8545},
doi = {https://doi.org/10.1016/j.ccr.2020.213466},
author = {Patric Zimmermann and Sergey Peredkov and Paula Macarena Abdala and Serena DeBeer and Moniek Tromp and Christoph Müller and Jeroen A. {van Bokhoven}},
}

@article {Platzman:Rydberg:1962,
    title = "Superexcited States of Molecules",
    author = "Robert L. Platzman",
    journal = "Radiat. Res.",
    year = "1962",
    month = sep,
    publisher = "Academic Press, Inc.",
    issn = "0033-7587",
    volume = "17",
    number = "3",
    pages = "419--425",
    doi = "10.2307/3571104",
}

@article{Klinker:Rydberg:2018,
    author = {Klinker, Markus and Marante, Carlos and Argenti, Luca and González-Vázquez, Jesús and Martín, Fernando},
    title = {Electron Correlation in the Ionization Continuum of
Molecules: Photoionization of {N$_2$} in the Vicinity of the
{Hopfield} Series of Autoionizing States},
    journal = {J. Phys. Chem. Lett.},
    volume = {9},
    number = {4},
    pages = {756-762},
    year = {2018},
    month = {01},
    issn = {1948-7185},
    doi = {10.1021/acs.jpclett.7b03220},
}

@article{Plunkett:Rydberg:2019,
  title={Ultrafast {Rydberg-state} dissociation in oxygen: Identifying the role of multielectron excitations},
  author={Plunkett, Alexander and Harkema, Nathan and Lucchese, Robert R and McCurdy, C William and Sandhu, Arvinder},
  journal={Phys. Rev. A},
  volume={99},
  number={6},
  pages={063403},
  year={2019},
  publisher={APS}
}

@article{Rabadan:Rydberg:2021,
  title={Toward a description of electron-induced dissociative excitation in {H$_2$O$^+$}: Investigation of three resonances above the {$\tilde{B}$} state},
  author={Rabad{\'a}n, I and Gorfinkiel, JD},
  journal={Phys. Rev. A},
  volume={103},
  number={3},
  pages={032804},
  year={2021},
  publisher={APS}
}

@article{Scrinzi:Stark:2006,
  title={Attosecond physics},
  author={Scrinzi, Armin and Ivanov, M Yu and Kienberger, Reinhard and Villeneuve, David M},
  journal={J. Phys. B: At., Mol. Opt. Phys.},
  volume={39},
  number={1},
  pages={R1--R37},
  year={2006}
}

@article{Gallmann:Stark:2012,
  title={Attosecond science: recent highlights and future trends},
  author={Gallmann, Lukas and Cirelli, Claudio and Keller, Ursula},
  journal={Annu. Rev. Phys. Chem.},
  volume={63},
  number={1},
  pages={447--469},
  year={2012},
  publisher={Annual Reviews}
}

@article{Cederbaum:ICD:1997,
  title={Giant intermolecular decay and fragmentation of clusters},
  author={Cederbaum, LS and Zobeley, J and Tarantelli, Francesco},
  journal={Phys. Rev. Lett.},
  volume={79},
  number={24},
  pages={4778},
  year={1997},
  publisher={APS}
}

@article{Jahnke:ICD:2020,
  title={Interatomic and intermolecular {C}oulombic decay},
  author={Jahnke, Till and Hergenhahn, Uwe and Winter, Bernd and Dörner, Reinhard and Frühling, Ulrike and Demekhin, Philipp V and Gokhberg, Kirill and Cederbaum, Lorenz S and Ehresmann, Arno and Knie, Andr{\'e} and others},
  journal={Chem. Rev.},
  volume={120},
  number={20},
  pages={11295--11369},
  year={2020},
  publisher={ACS Publications}
}

@article{Ouchi:ICD:2011,
  title={Three-electron interatomic {C}oulombic decay from the inner-valence double-vacancy states in {NeAr}},
  author={Ouchi, T and Sakai, K and Fukuzawa, H and Liu, X-J and Higuchi, I and Tamenori, Y and Nagaya, K and Iwayama, H and Yao, M and Zhang, D and others},
  journal={Phys. Rev. Lett.},
  volume={107},
  number={5},
  pages={053401},
  year={2011},
  publisher={APS}
}

@book{Ingolfsson:DEA:2019,
  title={Low-Energy Electrons: Fundamentals and Applications},
  author={Ing{\'o}lfsson, Oddur},
  year={2019},
  publisher={CRC Press}
}

@incollection{Fabrikant:DEA:2017,
title = {Chapter Nine - Recent Progress in Dissociative Electron Attachment: From Diatomics to Biomolecules},
editor = {Ennio Arimondo and Chun C. Lin and Susanne F. Yelin},
series = {Advances In Atomic, Molecular, and Optical Physics},
publisher = {Academic Press},
volume = {66},
pages = {545-657},
year = {2017},
issn = {1049-250X},
doi = {https://doi.org/10.1016/bs.aamop.2017.02.002},
author = {Ilya I. Fabrikant and Samuel Eden and Nigel J. Mason and Juraj Fedor}
}

@article{Aslam:Plasmon:2018,
  title={Catalytic conversion of solar to chemical energy on plasmonic metal nanostructures},
  author={Aslam, Umar and Rao, Vishal Govind and Chavez, Steven and Linic, Suljo},
  journal={Nat. Catal.},
  volume={1},
  number={9},
  pages={656--665},
  year={2018},
  publisher={Nature Publishing Group UK London}
}

@article{Martirez:Plasmon:2021,
   author = "Martirez, John Mark P. and Bao, Junwei Lucas and Carter, Emily A.",
   title = "First-Principles Insights into Plasmon-Induced Catalysis", 
   journal= "Annu. Rev. Phys. Chem.",
   year = "2021",
   volume = "72",
   number = "Volume 72, 2021",
   pages = "99-119",
   doi = "https://doi.org/10.1146/annurev-physchem-061020-053501",
   publisher = "Annual Reviews",
   issn = "1545-1593",
   type = "Journal Article"
  }

@article{Boudaiffa:Damage:2000,
  title={Resonant formation of {DNA} strand breaks by low-energy (3 to 20 e{V}) electrons},
  author={Bouda{\" i}ffa, Badia and Cloutier, Pierre and Hunting, Darel and Huels, Michael A and Sanche, L{\'e}on},
  journal={Science},
  volume={287},
  number={5458},
  pages={1658--1660},
  year={2000},
  publisher={American Association for the Advancement of Science}
}

@article{Gorfinkiel:Review:2020,
	author = {Gorfinkiel, Jimena D. },
	doi = {10.1140/epjd/e2020-100550-7},
	journal = {Eur. Phys. J. D},
	number = {3},
	pages = {51},
	title = {Electron collisions with molecules and molecular clusters},
	volume = {74},
	year = {2020},
}

@Book{Moiseyev:Book:11, 
  author =       {N. Moiseyev},
  ALTeditor =    {},
  title =        {Non-Hermitian Quantum Mechanics},
  publisher =    {Cambridge University Press},
  year =         {2011}
}

@article{Jagau:Review:2022,
    author = {Jagau, Thomas-C.},
    title = {Theory of electronic resonances: fundamental aspects and recent advances},
    journal = {Chem. Commun.},
    volume = {58},
    number = {34},
    pages = {5205-5224},
    year = {2022},
    month = {04},
    issn = {1359-7345},
    doi = {10.1039/d1cc07090h},
    url = {https://doi.org/10.1039/d1cc07090h}
}

@article{Aguilar:CS:1971,
author = {J. Aguilar and J. M. Combes},
journal = cmp,
year = {1971},
pages = {269-279},
volume = {22},
title = {A class of analytic perturbations for one-body {Schr{\"o}dinger Hamiltonians}},
}

@article{Balslev:CS:1971,
author = {E. Balslev and J. M. Combes},
title = {Spectral properties of many-body {Schr{\"o}dinger}
operators with dilatation-analytic interactions},
journal = cmp,
year = {1971},
pages = {280-294},
volume = {22},
}

@article{Simon:CS:1972,
author = {B. Simon},
title = {Quadratic form techniques and the {Balslev-Combes} theorem},
pages =  {1-9}, 
year = {1972},
journal = cmp,
volume = {27},
}

@Article{Moiseyev:CSReview:1998,
  author =       {N. Moiseyev},
  title =        {Quantum theory of resonances: {Calculating} energies, widths and cross-sections by complex scaling},
  journal =      prp,
  year =         {1998},
  volume =       {302},
  pages =        {212-293},
}

@Article{McCurdy:BO:1978,
  author =       {C. W. McCurdy and T. N. Rescigno},
  title =        {Extension of the Method of Complex Basis Functions to Molecular Resonances},
  journal =      prl,
  year =         {1978},
  volume =       {41},
  pages =        {1364-1368},
}

@Article{Moiseyev:CSH2:1979, 
  author =       {N. Moiseyev and C. Corcoran},
  title =        {Autoionizing states of {H}$_2$ and {H}$_2^-$ using the complex-scaling method},
  journal =      pra,
  year =         {1979},
  volume =       {20},
  pages =        {814-817},
}

@article{Austin:CAP:1985,
author = {G. Jolicard and E. J. Austin},
journal = cpl,
volume = {121},
pages = {106-110},
title = {Optical potential stabilisation method for predicting resonance levels},
year = {1985} 
}

@article{Meyer:CAP:1993,
author = {U. V. Riss and H.-D. Meyer}, 
journal = jpb,   
year = {1993},   
volume = {26},   
pages = {4503-4536},
title = {Calculation of resonance energies and widths using the
complex absorbing potential method},
}

@Article{Jagau:Review:2017,
  author =       {T.-C. Jagau and K. B. Bravaya and A. I. Krylov},
  title =        {Extending Quantum Chemistry of Bound States to Electronic Resonances},
  journal =      arpc,
  year =         {2017},
  volume =       {68},
  pages =        {525-553},
  OPTnote =      {doi: 10.1146/annurev-physchem-052516-050622},
}

@Article{Zuev:CAP:14,
  author =       {D. Zuev and  T.-C. Jagau and K. B. Bravaya 
and E. Epifanovsky and Y. Shao and E. Sundstrom and M. Head-Gordon
and A. I. Krylov},
  title =        {Complex absorbing potentials within {EOM-CC} family of methods:
{Theory}, Implementation, and Benchmarks},
  journal =      jcp,
  year =         {2014},
  volume =       {141},
  number =       {2},
  pages =        {024102},
  eid = {024102} 
}

@Article{Jagau:CAP:13,
  author =       {T.-C. Jagau and D. Zuev and  K. B. Bravaya and E. Epifanovsky and A. I. Krylov},
  title =        {A fresh look at resonances and complex absorbing potentials:
{Density}  matrix based approach},
  journal =      jpcl,
  year =         {2014},
  volume =       {5},
  pages =        {310--315}
}

@Article{Jagau:PES:14,
  author =       {T.-C. Jagau and A. I. Krylov},
  title =        {Complex absorbing potential equation-of-motion coupled-cluster method yields smooth and internally consistent potential energy surfaces and lifetimes for molecular resonances},
  journal =      jpcl,
  year =         {2014},
  volume =       {5},
  pages =        {3078-3085}
}

@article{Matz:Auger:2022,
    author = {Matz, Florian and Jagau, Thomas-C.},
    title = {Molecular {A}uger decay rates from complex-variable coupled-cluster theory},
    journal = jcp,
    volume = {156},
    number = {11},
    pages = {114117},
    year = {2022},
    month = {03},
    issn = {0021-9606},
    doi = {10.1063/5.0075646},
    url = {https://doi.org/10.1063/5.0075646}
}

@article{Utku:RICC2:2026,
    author = {Utku, Cansu and Paran, Garrette Pauley and Jagau, Thomas-C.},
    title = {Complex-energy second-order approximate coupled-cluster methods for electronic resonances},
    journal = jcp,
    volume = {163},
    number = {24},
    pages = {244103},
    year = {2025},
    month = {12},
    issn = {0021-9606},
    doi = {10.1063/5.0305953},
    url = {https://doi.org/10.1063/5.0305953}
}

@article{Moorby:Swave:2024,
    author = {Moorby, Robin E. and Parravicini, Valentina and Alessio, Maristella and Jagau, Thomas-C.},
    title = {Signatures of s-wave scattering in bound electronic states},
    journal = pccp,
    volume = {26},
    number = {8},
    pages = {6532-6539},
    year = {2024},
    month = {02},
    issn = {1463-9076},
    doi = {10.1039/d4cp00181h},
    url = {https://doi.org/10.1039/d4cp00181h}
}

@article{Moorby:Swave:2026,
    author = {Moorby, Robin E. and Mukherjee, Madhubani and Utku, Cansu and Jagau, Thomas-C. and Krylov, Anna I.},
    title = {A Fresh Look at Signatures of s-Wave Scattering: Symmetry and the Breakdown of the {Born–Oppenheimer} Approximation},
    journal = jpca,
    volume = {130},
    number = {25},
    pages = {4745-4758},
    year = {2026},
    month = {06},
    doi = {10.1021/acs.jpca.6c00547},
    url = {https://doi.org/10.1021/acs.jpca.6c00547}
}

@article{White:CBFSCF:2016,
    author = {White, Alec F. and McCurdy, C. William and Head-Gordon, Martin},
    title = {Restricted and unrestricted non-{H}ermitian {H}artree-{F}ock: Theory, practical considerations, and applications to metastable molecular anions},
    journal = jcp,
    volume = {143},
    number = {7},
    pages = {074103},
    year = {2015},
    month = {08},
    issn = {0021-9606},
    doi = {10.1063/1.4928529}
}

@article{White:CBFCCSD:2017,
    author = {White, Alec F. and Epifanovsky, Evgeny and McCurdy, C. William and Head-Gordon, Martin},
    title = {Second order {M}øller-{P}lesset and coupled cluster singles and doubles methods with complex basis functions for resonances in electron-molecule scattering},
    journal = jcp,
    volume = {146},
    number = {23},
    pages = {234107},
    year = {2017},
    month = {06},
    issn = {0021-9606},
    doi = {10.1063/1.4986950}
}

@article{Pulay:HFGradient:1969,
author = {P. Pulay},
title = {Ab initio calculation of force constants and equilibrium geometries in polyatomic molecules},
journal = mp,
volume = {17},
number = {2},
pages = {197-204},
year = {1969},
doi = {10.1080/00268976900100941},
URL = {https://doi.org/10.1080/00268976900100941}
}

@incollection{Helgaker:Gradient:1988,
title = {Analytical Calculation of Geometrical Derivatives in Molecular Electronic Structure Theory},
editor = {Per-Olov Löwdin},
booktitle = {Advances in Quantum Chemistry},
publisher = {Academic Press},
volume = {19},
pages = {183-245},
year = {1988},
issn = {0065-3276},
author = {Trygve Helgaker and Poul Jørgensen},
}

@article{Jorgensen:MP2Gradient:1988,
    author = {Jørgensen, Poul and Helgaker, Trygve},
    title = {Møller–{P}lesset energy derivatives},
    journal = jcp,
    volume = {89},
    number = {3},
    pages = {1560-1570},
    year = {1988},
    month = {08},
    issn = {0021-9606},
    doi = {10.1063/1.455152}
}

@article{Scheiner:CCSDGradient:1987,
    author = {Scheiner, Andrew C. and Scuseria, Gustavo E. and Rice, Julia E. and Lee, Timothy J. and {Schaefer III}, Henry F.},
    title = {Analytic evaluation of energy gradients for the single and double excitation coupled cluster ({CCSD}) wave function: Theory and application},
    journal = jcp,
    volume = {87},
    number = {9},
    pages = {5361-5373},
    year = {1987},
    month = {11},
    issn = {0021-9606},
    doi = {10.1063/1.453655},
    url = {https://doi.org/10.1063/1.453655}
}

@article{Gauss:CCGradient:1991,
    author = {Gauss, Jürgen and Stanton, John F. and Bartlett, Rodney J.},
    title = {Coupled‐cluster open‐shell analytic gradients: Implementation of the direct product decomposition approach in energy gradient calculations},
    journal = jcp,
    volume = {95},
    number = {4},
    pages = {2623-2638},
    year = {1991},
    month = {08},
    issn = {0021-9606},
    doi = {10.1063/1.460915},
    url = {https://doi.org/10.1063/1.460915}
}

@article{Stanton:EOMCCGradient:1993,
    author = {Stanton, John F.},
    title = {Many‐body methods for excited state potential energy surfaces. {I}. {G}eneral theory of energy gradients for the equation‐of‐motion coupled‐cluster method},
    journal = jcp,
    volume = {99},
    number = {11},
    pages = {8840-8847},
    year = {1993},
    month = {12},
    issn = {0021-9606},
    doi = {10.1063/1.465552},
    url = {https://doi.org/10.1063/1.465552}
}

@article{Stanton:EOMIPGradient:1994,
    author = {Stanton, John F. and Gauss, Jürgen},
    title = {Analytic energy derivatives for ionized states described by the equation‐of‐motion coupled cluster method},
    journal = jcp,
    volume = {101},
    number = {10},
    pages = {8938-8944},
    year = {1994},
    month = {11},
    issn = {0021-9606},
    doi = {10.1063/1.468022}
}

@article{Benda:Gradient:2017,
    author = {Benda, Zsuzsanna and Jagau, Thomas-C.},
    title = {Communication: Analytic gradients for the complex absorbing potential equation-of-motion coupled-cluster method},
    journal = jcp,
    volume = {146},
    number = {3},
    pages = {031101},
    year = {2017},
    month = {01},
    issn = {0021-9606},
    doi = {10.1063/1.4974094},
    url = {https://doi.org/10.1063/1.4974094}
}

@article{Benda:Optimization:2018,
    author = {Benda, Zsuzsanna and Rickmeyer, Kerstin and Jagau, Thomas-C.},
    title = {Structure Optimization of Temporary Anions},
    journal = jctc,
    volume = {14},
    number = {7},
    pages = {3468-3478},
    year = {2018},
    month = {06},
    issn = {1549-9618},
    doi = {10.1021/acs.jctc.8b00128},
    url = {https://doi.org/10.1021/acs.jctc.8b00128}
}

@article{Mondal:ProjectedCAPGradient:2025,
    author = {Mondal, Soubhik and Bravaya, Ksenia B.},
    title = {Analytic Nuclear
Gradients for Complex Potential Energy
Surfaces: A Projected {CAP} Approach},
    journal = jctc,
    volume = {21},
    number = {12},
    pages = {5986-5996},
    year = {2025},
    month = {06},
    issn = {1549-9618},
    doi = {10.1021/acs.jctc.5c00418}
}

@incollection{Gauss:Properties:2000,
  author    = {Gauss, J{\"u}rgen},
  title     = {Molecular Properties},
  booktitle = {Modern Methods and Algorithms of Quantum Chemistry},
  editor    = {Grotendorst, Johannes},
  edition   = {2},
  series    = {NIC Series},
  volume    = {3},
  pages     = {541--592},
  publisher = {John von Neumann Institute for Computing},
  address   = {J{\"u}lich, Germany},
  year      = {2000},
  isbn      = {3-00-005834-6},
  url       = {https://juser.fz-juelich.de/record/152462}
}

@article{Burrow:CH2OAnion:1976,
author = {Burrow, P. D. and Michejda, J. A.},
title = {Electron transmission study of the formaldehyde electron affinity},
journal = cpl,
volume = {42},
number = {2},
pages = {223-226},
year = {1976},
issn = {0009-2614},
doi = {https://doi.org/10.1016/0009-2614(76)80351-X}
}

@article{Aflatooni:HCOOHAnion:2001,
    author = {Aflatooni, K. and Hitt, B. and Gallup, G. A. and Burrow, P. D.},
    title = {Temporary anion states of selected amino acids},
    journal = jcp,
    volume = {115},
    number = {14},
    pages = {6489-6494},
    year = {2001},
    month = {10},
    issn = {0021-9606},
    doi = {10.1063/1.1404147},
    url = {https://doi.org/10.1063/1.1404147}
}

@inproceedings{Tronc:HCOOH:1987,
  author    = {M. Tronc and M. Allan and F. Edard},
  booktitle = {Abstract of Contributed Papers, XVth ICPEAC},
  year      = {1987},
  address   = {Amsterdam},
  publisher = {North-Holland},
  pages     = {335}
}

@article{Sanche:C2H4Anion:1973,
    author = {Sanche, L. and Schulz, G. J.},
    title = {Electron transmission spectroscopy: Resonances in triatomic molecules and hydrocarbons},
    journal = jcp,
    volume = {58},
    number = {2},
    pages = {479-493},
    year = {1973},
    month = {01},
    issn = {0021-9606},
    doi = {10.1063/1.1679228},
    url = {https://doi.org/10.1063/1.1679228}
}

@article{Burrow:C2H4Anion:1975,
title = {On the electron affinities of ethylene and 1,3-butadiene},
journal = cpl,
volume = {36},
number = {5},
pages = {594-598},
year = {1975},
issn = {0009-2614},
doi = {https://doi.org/10.1016/0009-2614(75)85346-2},
url = {https://www.sciencedirect.com/science/article/pii/0009261475853462},
author = {Burrow, P. D. and Jordan, K. D.}
}

@article{Sankari:WaterAuger:2003,
title = {Vibrationally resolved {O} 1s photoelectron spectrum of water},
journal = cpl,
volume = {380},
number = {5},
pages = {647-653},
year = {2003},
issn = {0009-2614},
doi = {https://doi.org/10.1016/j.cplett.2003.08.108},
url = {https://www.sciencedirect.com/science/article/pii/S0009261403015720},
author = {R. Sankari and M. Ehara and H. Nakatsuji and Y. Senba and K. Hosokawa and H. Yoshida and A. {De Fanis} and Y. Tamenori and S. Aksela and K. Ueda}
}

@article{Hitchcock:C2H4Auger:1977,
title = {Carbon {K}-shell excitation of {C$_2$H$_2$}, {C$_2$H$_4$}, {C$_2$H$_6$} and {C$_6$H$_6$} by 2.5 ke{V} electron impact},
journal = {J. Electron Spectrosc. Relat. Phenom.},
volume = {10},
number = {3},
pages = {317-330},
year = {1977},
issn = {0368-2048},
doi = {https://doi.org/10.1016/0368-2048(77)85029-9},
url = {https://www.sciencedirect.com/science/article/pii/0368204877850299},
author = {A.P. Hitchcock and C.E. Brion}
}

@article{Kuramoto:CH2OAuger:2005,
title = {{C} 1s and {O} 1s photoelectron spectra of formaldehyde with satellites: theory and experiment},
journal = {J. Electron Spectrosc. Relat. Phenom.},
volume = {142},
number = {3},
pages = {253-259},
year = {2005},
issn = {0368-2048},
doi = {https://doi.org/10.1016/j.elspec.2004.09.011},
url = {https://www.sciencedirect.com/science/article/pii/S036820480400369X},
author = {K. Kuramoto and M. Ehara and H. Nakatsuji and M. Kitajima and H. Tanaka and A. {De Fanis} and Y. Tamenori and K. Ueda}
}

@article{DeFanis:CH2OAuger:2005,
  title = {Effect of vibrations on {C} $1s$ photoemission in formaldehyde in the shape resonance region},
  author = {De Fanis, A. and Mistrov, D. A. and Kitajima, M. and Hoshino, M. and Shindo, H. and Tanaka, T. and Tanaka, H. and Tamenori, Y. and Pavlychev, A. A. and Ueda, K.},
  journal = {Phys. Rev. A},
  volume = {71},
  issue = {5},
  pages = {052510},
  numpages = {7},
  year = {2005},
  month = {May},
  publisher = {American Physical Society},
  doi = {10.1103/PhysRevA.71.052510}
}

@article{Kempgens:C2H4Auger:1998,
title = {Threshold phenomena in high-resolution core-level photoelectron spectroscopy: the ethene molecule},
journal = {J. Electron Spectrosc. Relat. Phenom.},
volume = {93},
number = {1},
pages = {39-48},
year = {1998},
issn = {0368-2048},
doi = {https://doi.org/10.1016/S0368-2048(98)00156-X},
url = {https://www.sciencedirect.com/science/article/pii/S036820489800156X},
author = {B. Kempgens and A. Kivimäki and B.S. Itchkawitz and H.M. Köppe and M. Schmidbauer and M. Neeb and K. Maier and J. Feldhaus and A.M. Bradshaw}
}

@article{Boys:BoysFunction:1950,
 ISSN = {00804630},
 author = {S. F. Boys},
 journal = {Proc. R. Soc. London, Ser. A},
 number = {1063},
 pages = {542--554},
 publisher = {The Royal Society},
 title = {Electronic Wave Functions. {I}. {A} General Method of Calculation for the Stationary States of Any Molecular System},
 urldate = {2026-10-20},
 volume = {200},
 year = {1950}
}

@article{Obara:TEIs:1986,
    author = {Obara, S. and Saika, A.},
    title = {Efficient recursive computation of molecular integrals over {C}artesian {G}aussian functions},
    journal = jcp,
    volume = {84},
    number = {7},
    pages = {3963-3974},
    year = {1986},
    month = {04},
    issn = {0021-9606},
    doi = {10.1063/1.450106}
}

@article{McMurchie:TEIs:1978,
title = {One- and two-electron integrals over {C}artesian {G}aussian functions},
journal = {J. Comp. Phys.},
volume = {26},
number = {2},
pages = {218-231},
year = {1978},
issn = {0021-9991},
doi = {https://doi.org/10.1016/0021-9991(78)90092-X},
author = {Larry E McMurchie and Ernest R Davidson}
}

@article{HeadGordon:TEIs:1988,
    author = {Head‐Gordon, Martin and Pople, John A.},
    title = {A method for two‐electron {Gaussian} integral and integral derivative evaluation using recurrence relations},
    journal = jcp,
    volume = {89},
    number = {9},
    pages = {5777-5786},
    year = {1988},
    month = {11},
    doi = {10.1063/1.455553},
    url = {https://doi.org/10.1063/1.455553}
}

@article{White:CBF:2015,
    author = {White, Alec F. and Head-Gordon, Martin and McCurdy, C. William},
    title = {Complex basis functions revisited: Implementation with applications to carbon tetrafluoride and aromatic {N}-containing heterocycles within the static-exchange approximation},
    journal = jcp,
    volume = {142},
    number = {5},
    pages = {054103},
    year = {2015},
    month = {02},
    issn = {0021-9606},
    doi = {10.1063/1.4906940},
    url = {https://doi.org/10.1063/1.4906940}
}

@article{Taketa:Integrals:1966,
author = {Taketa ,Hiroshi and Huzinaga ,Sigeru and O-ohata ,Kiyosi},
title = {Gaussian-Expansion Methods for Molecular Integrals},
journal = {J. Phys. Soc. Jpn.},
volume = {21},
number = {11},
pages = {2313-2324},
year = {1966},
doi = {10.1143/JPSJ.21.2313}
}

@article{Petersson:Integrals:2009,
doi = {10.1088/0143-0807/31/1/004},
url = {https://doi.org/10.1088/0143-0807/31/1/004},
year = {2009},
month = {nov},
publisher = {},
volume = {31},
number = {1},
pages = {37},
author = {Petersson, T and Hellsing, B},
title = {A detailed derivation of {Gaussian} orbital-based matrix elements in electron structure calculations},
journal = {Eur. J. Phys.}
}

@article{Gao:Integrals:2011,
author = {Gao, Bin and Thorvaldsen, Andreas J. and Ruud, Kenneth},
title = {{GEN1INT}: A unified procedure for the evaluation of one-electron integrals over {Gaussian} basis functions and their geometric derivatives},
journal = ijqc,
volume = {111},
number = {4},
pages = {858-872},
doi = {https://doi.org/10.1002/qua.22886},
url = {https://onlinelibrary.wiley.com/doi/abs/10.1002/qua.22886},
year = {2011}
}

@article{QChem,
author = {E. Epifanovsky and A. T. B. Gilbert and X. Feng and J. Lee and Y. Mao and N. Mardirossian and P. Pokhilko and A. F. White and M. P. Coons and A. L. Dempwolff and Z. Gan and D. Hait and P. R. Horn and L. D. Jacobson and I. Kaliman and J. Kussmann and A. W. Lange and K. U. Lao and D. S. Levine and J. Liu and S. C. McKenzie  and A. F. Morrison and K. D. Nanda and F. Plasser and D. R. Rehn and M. L. Vidal and Z.-Q. You and Y. Zhu and B. Alam and B. J. Albrecht and A. Aldossary and E. Alguire and J. H. Andersen and V. Athavale and D. Barton and K. Begam and A. Behn and N. Bellonzi and Y. A. Bernard and E. J. Berquist and H. G. A. Burton and A. Carreras and K. Carter-Fenk and R. Chakraborty and A. D. Chien and K. D. Closser and V. Cofer-Shabica and S. Dasgupta and M. {de Wergifosse} and J. Deng and M. Diedenhofen and H. Do and S. Ehlert and P.-T. Fang and S. Fatehi and Q. Feng and T. Friedhoff and J. Gayvert and Q. Ge and G. Gidofalvi and M. Goldey and J. Gomes and C. E. Gonz{\'a}lez-Espinoza and S. Gulania  and A. O. Gunina and M. W. D. Hanson-Heine and P. H. P. Harbach and A. Hauser and M. F. Herbst and M. {Hern{\'a}ndez Vera} and M. Hodecker and Z. C. Holden and S. Houck and X. Huang and K. Hui and B. C. Huynh and M. Ivanov and {\'A}. J{\'a}sz and H. Ji and H. Jiang and B. Kaduk and S. K{\"a}hler and K. Khistyaev and J. Kim and G. Kis and P. Klunzinger and Z. Koczor-Benda and J. H. Koh and D. Kosenkov and L. Koulias and T. Kowalczyk and C. M. Krauter and K. Kue and A. Kunitsa and T. Kus  and I. Ladj{\'a}nszki and A. Landau  and K. V. Lawler and D. Lefrancois and S. Lehtola  and R. R. Li  and Y.-P. Li and J. Liang  and M. Liebenthal and H.-H. Lin  and Y.-S. Lin and F. Liu   and K.-Y. Liu and M. Loipersberger and A. Luenser and A. Manjanath and P. Manohar and E. Mansoor and S. F. Manzer and S.-P. Mao and A. V. Marenich and T. Markovich and S. Mason and S. A. Maurer  and P. F. McLaughlin and M. F. S. J. Menger and J.-M. Mewes and S. A. Mewes and P. Morgante and J. W. Mullinax and K. J. Oosterbaan and G. Paran and A. C. Paul and S. K. Paul and F. Pavo{\v s}evi{\'c} and Z. Pei and S. Prager and E. I. Proynov and {\' A}. R{\'a}k and E. Ramos-Cordoba and B. Rana and A. E. Rask  and A. Rettig and R. M. Richard and F. Rob and E. Rossomme and T. Scheele and M. Scheurer and M. Schneider and N. Sergueev and S. M. Sharada and W. Skomorowski and D. W. Small and C. J. Stein and Y.-C. Su and E. J. Sundstrom and Z. Tao and J. Thirman and G. J. Tornai and T. Tsuchimochi and N. M. Tubman and S. P. Veccham and O. Vydrov and J. Wenzel and J. Witte and A. Yamada and K. Yao and S. Yeganeh and S. R. Yost and A. Zech and I. Y. Zhang and X. Zhang and Y. Zhang and D. Zuev and A. Aspuru-Guzik and A. T. Bell and N. A. Besley and K. B. Bravaya  and B. R. Brooks and D. Casanova and J.-D. Chai and S. Coriani and C. J. Cramer  and G. Cserey and A. E. DePrince  and R. A. DiStasio and A. Dreuw  and B. D. Dunietz and T. R. Furlani and W. A. Goddard and S. Hammes-Schiffer and T. Head-Gordon and W. J. Hehre  and C.-P. Hsu and T.-C. Jagau  and Y. Jung  and A. Klamt and J. Kong and D. S. Lambrecht and W. Liang and N. J. Mayhall and C. W. McCurdy and J. B. Neaton and C. Ochsenfeld and J. A. Parkhill and R. Peverati and V. A. Rassolov and Y. Shao and L. V. Slipchenko and T. Stauch and R. P. Steele and J. E. Subotnik and A. J. W. Thom and A. Tkatchenko and D. G. Truhlar and T. {Van Voorhis} and T. A. Wesolowski and K. B. Whaley and H. L. Woodcock  and P. M. Zimmerman and S. Faraji and P. M. W. Gill and M. Head-Gordon and J. M. Herbert and A. I. Krylov},
title = {Software for the frontiers of quantum chemistry: {An} overview of developments in the {Q-Chem} 5 package},
journal = jcp,
volume = {155},
number = {8},
pages = {084801},
year = {2021},
doi = {10.1063/5.0055522},
}

@article{Epifanovsky:libtensor:2013,
author = {Epifanovsky, Evgeny and Wormit, Michael and Kuś, Tomasz and Landau, Arie and Zuev, Dmitry and Khistyaev, Kirill and Manohar, Prashant and Kaliman, Ilya and Dreuw, Andreas and Krylov, Anna I.},
title = {New implementation of high-level correlated methods using a general block tensor library for high-performance electronic structure calculations},
journal = jcc,
volume = {34},
number = {26},
pages = {2293-2309},
doi = {https://doi.org/10.1002/jcc.23377},
year = {2013}
}

@article{CFOUR,
    author = {Matthews, Devin A. and Cheng, Lan and Harding, Michael E. and Lipparini, Filippo and Stopkowicz, Stella and Jagau, Thomas-C. and Szalay, P{\' e}ter G. and Gauss, Jürgen and Stanton, John F.},
    title = {Coupled-cluster techniques for computational chemistry: The {CFOUR} program package},
    journal = jcp,
    volume = {152},
    number = {21},
    pages = {214108},
    year = {2020},
    month = {06},
    issn = {0021-9606},
    doi = {10.1063/5.0004837},
    url = {https://doi.org/10.1063/5.0004837}
}

@misc{gamma:vargamma,
note = {We highlight the use of upright $\Gamma$ for the effective two-particle density matrix in this expression to distinguish it from the decay width, $\varGamma$.}
}

@article{Dunning:pVTZ:1989,
    author = {Dunning, Thom H.},
    title = {Gaussian basis sets for use in correlated molecular calculations. {I}. {T}he atoms boron through neon and hydrogen},
    journal = jcp,
    volume = {90},
    pages = {1007-1023},
    year = {1989},
    doi = {10.1063/1.456153}
}

@article{Kendall:aug:1992,
    author = {Kendall, Rick A. and Dunning, Thom H. and Harrison, Robert J.},
    title = {Electron affinities of the first-row atoms revisited. {S}ystematic basis sets and wave functions},
    journal = jcp,
    volume = {96},
    pages = {6796-6806},
    year = {1992},
    doi = {10.1063/1.462569}
}

@article{Woon:PolarizedCore:1995,
    author = {Woon, David E. and Dunning, Thom H.},
    title = {Gaussian basis sets for use in correlated molecular calculations. {V}. {C}ore-valence basis sets for boron through neon},
    journal = jcp,
    volume = {103},
    pages = {4572-4585},
    year = {1995},
    doi = {10.1063/1.470645}
}

@article{Benda:MECPs:2018,
    author = {Benda, Zsuzsanna and Jagau, Thomas-C.},
    title = {Understanding Processes Following Resonant Electron
Attachment: Minimum-Energy Crossing Points between Anionic and Neutral
Potential Energy Surfaces},
    journal = jctc,
    volume = {14},
    number = {8},
    pages = {4216-4223},
    year = {2018},
    month = {08},
    issn = {1549-9618},
    doi = {10.1021/acs.jctc.8b00444}
}

@article{Benda:EPs:2018,
    author = {Benda, Zsuzsanna and Jagau, Thomas-C.},
    title = {Locating Exceptional Points on Multidimensional Complex-Valued
Potential Energy Surfaces},
    journal = jpcl,
    volume = {9},
    number = {24},
    pages = {6978-6984},
    year = {2018},
    month = {12},
    issn = {1948-7185},
    doi = {10.1021/acs.jpclett.8b03228}
}

@article{Gyamfi:AIMD:2022,
    author = {Gyamfi, Jerryman A. and Jagau, Thomas-C.},
    title = {Ab Initio Molecular Dynamics of Temporary
Anions Using Complex Absorbing Potentials},
    journal = jpcl,
    volume = {13},
    number = {36},
    pages = {8477-8483},
    year = {2022},
    month = {09},
    issn = {1948-7185},
    doi = {10.1021/acs.jpclett.2c01969}
}

@article{Chatterjee:NACs:2023,
    author = {Chatterjee, Koushik and Koczor-Benda, Zsuzsanna and Feng, Xintian and Krylov, Anna I. and Jagau, Thomas-C.},
    title = {Analytic Evaluation
of Nonadiabatic Couplings within
the Complex Absorbing Potential Equation-of-Motion Coupled-Cluster
Method},
    journal = jctc,
    volume = {19},
    number = {17},
    pages = {5821-5834},
    year = {2023},
    month = {09},
    issn = {1549-9618},
    doi = {10.1021/acs.jctc.3c00514}
}

@article{Moiseyev:ResonanceProperties:1978,
author = {N. Moiseyev and P.R. Certain and F. Weinhold},
title = {Resonance properties of complex-rotated hamiltonians},
journal = {Mol. Phys.},
volume = {36},
number = {6},
pages = {1613--1630},
year = {1978},
publisher = {Taylor \& Francis},
doi = {10.1080/00268977800102631}
}

@article{Moiseyev:Forces:2017,
    author = {Moiseyev, Nimrod},
    title = {Forces on nuclei moving on autoionizing molecular potential energy surfaces},
    journal = jcp,
    volume = {146},
    number = {2},
    pages = {024101},
    year = {2017},
    month = {01},
    issn = {0021-9606},
    doi = {10.1063/1.4973559}
}

@article{Shirley:EquivalentCore:1972,
title = {The effect of atomic and extra-atomic relaxation on atomic binding energies},
journal = cpl,
volume = {16},
number = {2},
pages = {220-225},
year = {1972},
issn = {0009-2614},
doi = {https://doi.org/10.1016/0009-2614(72)80260-4},
author = {D.A. Shirley}
}

@article{Bravaya:ComplexScaling:2013,
    author = {Bravaya, Ksenia B. and Zuev, Dmitry and Epifanovsky, Evgeny and Krylov, Anna I.},
    title = {Complex-scaled equation-of-motion coupled-cluster method with single and double substitutions for autoionizing excited states: Theory, implementation, and examples},
    journal = jcp,
    volume = {138},
    number = {12},
    pages = {124106},
    year = {2013},
    month = {03},
    issn = {0021-9606},
    doi = {10.1063/1.4795750}
}

@article{Sommerfeld:CAPCI:2001,
author = {Sommerfeld, Thomas and Santra, Robin},
title = {Efficient method to perform {CAP/CI} calculations for temporary anions},
journal = ijqc,
volume = {82},
number = {5},
pages = {218-226},
doi = {https://doi.org/10.1002/qua.1042},
year = {2001}
}

@article{Santra:CAP:2001,
    author = {Santra, Robin and Cederbaum, Lorenz S.},
    title = {An efficient combination of computational techniques for investigating electronic resonance states in molecules},
    journal = jcp,
    volume = {115},
    number = {15},
    pages = {6853-6861},
    year = {2001},
    month = {10},
    issn = {0021-9606},
    doi = {10.1063/1.1405117}
}

@article{Sommerfeld:Voronoi:2015,
    author = {Sommerfeld, Thomas and Ehara, Masahiro},
    title = {Complex Absorbing Potentials with {V}oronoi Isosurfaces
Wrapping Perfectly around Molecules},
    journal = jctc,
    volume = {11},
    number = {10},
    pages = {4627-4633},
    year = {2015},
    month = {10},
    issn = {1549-9618},
    doi = {10.1021/acs.jctc.5b00465}
}

@article{Gayvert:CAP:2022,
    author = {Gayvert, James R. and Bravaya, Ksenia B.},
    title = {Projected {CAP-EOM-CCSD} method for electronic resonances},
    journal = jcp,
    volume = {156},
    number = {9},
    pages = {094108},
    year = {2022},
    month = {03},
    issn = {0021-9606},
    doi = {10.1063/5.0082739}
}

@article{Gyamfi:CAP:2024,
    author = {Gyamfi, Jerryman A. and Jagau, Thomas-C.},
    title = {A New Strategy
to Optimize Complex Absorbing Potentials
for the Computation of Resonance Energies and Widths},
    journal = jctc,
    volume = {20},
    number = {3},
    pages = {1096-1107},
    year = {2024},
    month = {02},
    issn = {1549-9618},
    doi = {10.1021/acs.jctc.3c01039}
}

@article{Sanov:CAP:2026,
    author = {Sanov, Andrei},
    title = {How to Choose a
Better Box: Complex Absorbing Potential
Optimization for Anionic Resonances},
    journal = jpca,
    volume = {130},
    number = {10},
    pages = {2112-2125},
    year = {2026},
    month = {03},
    issn = {1089-5639},
    doi = {10.1021/acs.jpca.5c08585}
}

@article{Nooijen:EOMEA:1995,
    author = {Nooijen, Marcel and Bartlett, Rodney J.},
    title = {Equation of motion coupled cluster method for electron attachment},
    journal = jcp,
    volume = {102},
    number = {9},
    pages = {3629-3647},
    year = {1995},
    month = {03},
    issn = {0021-9606},
    doi = {10.1063/1.468592}
}

@article{Cederbaum:AugerBroadening:1991,
    author = {Cederbaum, L. S. and Campos, P. and Tarantelli, F. and Sgamellotti, A.},
    title = {Band shape and vibrational structure in {A}uger spectra: Theory and application to carbon monoxide},
    journal = jcp,
    volume = {95},
    number = {9},
    pages = {6634-6644},
    year = {1991},
    month = {11},
    issn = {0021-9606},
    doi = {10.1063/1.461533}
}

@article{Cederbaum:CVS:1980,
  title = {Many-body theory of core holes},
  author = {Cederbaum, L. S. and Domcke, W. and Schirmer, J.},
  journal = {Phys. Rev. A},
  volume = {22},
  issue = {1},
  pages = {206--222},
  numpages = {0},
  year = {1980},
  month = {Jul},
  publisher = {American Physical Society},
  doi = {10.1103/PhysRevA.22.206},
  url = {https://link.aps.org/doi/10.1103/PhysRevA.22.206}
}

@article{Frati:CVSGradient:2019,
    author = {Frati, Federica and de Groot, Frank and Cerezo, Javier and Santoro, Fabrizio and Cheng, Lan and Faber, Rasmus and Coriani, Sonia},
    title = {Coupled cluster study of the {X}-ray absorption spectra of formaldehyde derivatives at the oxygen, carbon, and fluorine {K}-edges},
    journal = jcp,
    volume = {151},
    number = {6},
    pages = {064107},
    year = {2019},
    month = {08},
    issn = {0021-9606},
    doi = {10.1063/1.5097650}
}

@article{Camps:RICCSD:2025,
    author = {Camps, Simen and Utku, Cansu and Creutzberg, Joel and Jagau, Thomas-C.},
    title = {Complex-Variable
Equation-of-Motion Coupled-Cluster
Singles and Doubles Theory with the Resolution-of-the-Identity Approximation},
    journal = jpca,
    volume = {129},
    number = {21},
    pages = {4798-4811},
    year = {2025},
    month = {05},
    issn = {1089-5639},
    doi = {10.1021/acs.jpca.5c01313}
}

@article{Brian:DissociativeRecombination:1990,
title = {The dissociative recombination of molecular ions},
journal = {Phys. Rep.},
volume = {186},
number = {5},
pages = {215-248},
year = {1990},
issn = {0370-1573},
doi = {https://doi.org/10.1016/0370-1573(90)90159-Y},
author = {J. Brian and A. Mitchell}
}

@article{Bardsley:DissociativeRecombination:1968,
doi = {10.1088/0022-3700/1/3/304},
url = {https://doi.org/10.1088/0022-3700/1/3/304},
year = {1968},
month = {may},
volume = {1},
number = {3},
pages = {365},
author = {J N Bardsley},
title = {The theory of dissociative recombination},
journal = {J. Phys. B: At. Mol. Phys.}
}

@article{Coriani:CVS:2015,
    author = {Coriani, Sonia and Koch, Henrik},
    title = {Communication: X-ray absorption spectra and core-ionization potentials within a core-valence separated coupled cluster framework},
    journal = jcp,
    volume = {143},
    number = {18},
    pages = {181103},
    year = {2015},
    month = {11},
    issn = {0021-9606},
    doi = {10.1063/1.4935712}
}

@article{Stanton:FrequencyReview:2000,
author = {John F. Stanton and Jurgen Gauss},
title = {Analytic second derivatives in high-order many-body perturbation and coupled-cluster theories: Computational considerations and applications},
journal = {Int. Rev. Phys. Chem.},
volume = {19},
number = {1},
pages = {61--95},
year = {2000},
publisher = {Taylor \& Francis},
doi = {10.1080/014423500229864}
}

@mastersthesis{Schepens:LLLDecay:2026,
    author = {Schepens, Lucas},
    title = {Super {Coster}-{Kronig} decay in molecules exclusively composed of light elements},
    school = {KU Leuven},
    address = {Leuven, Belgium},
    year = {2026}
}

@article{Eland:LLLDecay:2008,
title = {Spectra of the dications of benzene, naphthalene and azulene},
journal = {Chem. Phys.},
volume = {345},
number = {1},
pages = {82-86},
year = {2008},
issn = {0301-0104},
doi = {https://doi.org/10.1016/j.chemphys.2008.01.037},
author = {John H.D. Eland}
}

@article{Molloy:LLLDecay:2006,
title = {Double photoionisation of toluene},
journal = cpl,
volume = {421},
number = {1},
pages = {31-35},
year = {2006},
issn = {0009-2614},
doi = {https://doi.org/10.1016/j.cplett.2006.01.051},
author = {Ryan D. Molloy and John H.D. Eland}
}

@article{Arvinder:LLLDecay:2008,
author = {Arvinder S. Sandhu  and Etienne Gagnon  and Robin Santra  and Vandana Sharma  and Wen Li  and Phay Ho  and Predrag Ranitovic  and C. Lewis Cocke  and Margaret M. Murnane  and Henry C. Kapteyn },
title = {Observing the Creation of Electronic {F}eshbach Resonances in Soft {X}-ray-Induced {O}$_2$ Dissociation},
journal = {Science},
volume = {322},
number = {5904},
pages = {1081-1085},
year = {2008},
doi = {10.1126/science.1164498}
}

@article{Idebohn:LLLDecay:2026,
	author = {Ideb{\"o}hn, Veronica Daver and Gloriod, Antoine and Squibb, Richard J. and Roos, Andreas Hult and Behera, Nihar Ranjan and Kanungo, Ishita and Gustafsson, Elias and G{\"a}llblad, Simon and Berglund, Saga and Olsson, Emelie and Al-Mogren, Muneerah Mogren and {\"O}hrwall, Gunnar and Nyman, Gunnar and Dyke, John M. and Eland, John H. D. and Hochlaf, Majdi and Feifel, Raimund},
	date = {2026/08/22},
	doi = {10.1038/s41598-026-66986-0},
	isbn = {2045-2322},
	journal = {Sci. Rep.},
	number = {1},
	pages = {26361},
	title = {Single-photon double ionization of ozone: Experiment and theory},
	volume = {16},
	year = {2026}
}

\end{document}